# Surface Electronic Structures and Field Emission Currents at Sodium Overlayers on Low-Index Tungsten Surfaces

Z.A. Ibrahim* and M.J.G. Lee†
(Department of Physics, University of Toronto, Toronto, Ontario, Canada)

The total energy distributions (TEDs) of the emission currents in field emission and surface photofield emission and the overlayer-induced modifications in the surface electronic structures from the technologically important W surfaces with the commensurate W(100)/Na c(2×2), W(110)/Na (2×2) and W(111)/Na (1×1) overlayers are calculated. The TEDs obtained by our recent numerical method that extends the full-potential linear augmented plane wave method for the electronic structures to the study of field and photofield emission are used to interpret the shifts of the peaks in the experimental TEDs in field emission and photofield emission from the W(100) and W(110) surfaces at sub-monolayer and monolayer Na coverage. Hybridization of the 3s Na states with the pairs of $d_z{}^2$-like surface states of the strong Swanson hump in clean W(100) and surface resonances in clean W(111) below the Fermi energy shifts these W states by about −1.2 eV and −1.0 eV, thus stabilizing these states, to yield new strong peaks in the TEDs in field emission and photofield emission from W(100)/Na c(2×2) and W(111)/Na (1×1) respectively. The effect of Na intralayer interactions are discussed and are shown to shift the strong *s*- and *p*-like peaks in the surface density of states of W(110) below and above the Fermi energy respectively to lower energy with increased Na coverage, in agreement with experiments.

## I. INTRODUCTION

Field emission has been extensively used in technological applications for imaging as well as to study the electronic properties at metal and semiconductor surfaces and thin film overlayers[1,2]. Modifications in the electron emission characteristics of transition metals due to the adsorption of alkali metal overlayers and their oxides is of great interest both fundamentally and technologically[3]. The reduction in the work function, induced by the overlayer, enhances the quantum mechanical tunneling of electrons through the surface making these interfaces suitable for thermionic cathode and photoelectron emission device applications. To understand and improve the performance of these devices, it is important to microscopically interpret the physical mechanisms at the interfaces that modify the electronic structures and emission current with changing coverage. Details of the theoretically extracted electronic structure are expected to shed light on the role played by surface states and surface resonances in the substrate - overlayer bonding, the nature of bonds formed, the charge shifts and modification of the potential induced at the interface by the overlayer. While several such overlayer systems have been extensively studied theoretically[4,5,6], the theoretical work on the tungsten-sodium (W-Na) interface has been limited to only the (100) plane[3], or was using simplified formalisms that neglect the dependence of the substrate electronic structure on surface orientation[7,8].

† Deceased 21 April 2009

Most experimental studies of the W-Na interfaces were focused on work function measurements and understanding the Na growth mechanism[3, 9-12]. Photoemission measurements[13] from W(110) covered by Na, K and Cs from low up to complete one monolayer coverage were used to analyze the binding energies of the alkali s, and tungsten 3p and 4f core electrons. Metastable Impact Electron Spectroscopy and Ultraviolet Photoemission Spectroscopy were used to study the electron spectra from Li, Na, K and Cs overlayers on W(110)[14]. An experimental investigation of the valence electronic structure of the W-Na-vacuum interfaces for W(100) and W(110) was carried out by Derraa and Lee[15] by the surface sensitive field emission (FE) and photofield emission (PFE) techniques. Na coverage was in the range from 0 to $\theta = 1/2$ and 0 to $\theta = 1/4$ respectively, where $\theta$ is the ratio between the number of atoms in the overlayer and that in the outermost substrate layer.

Although many experimental results on W-Na have been available for more than a decade and in spite of the technological importance of these interfaces, only a few theoretical investigations - including only one recent[3] parameter-free, *ab-initio* study- of the electronic structure of these interfaces are reported. The recent work[3] that investigated Na overlayer growth at W electrodes at different Na pressure in high pressure Na discharge lamps discusses the pressure dependent work function at the W(100) surface at different coverages. The measured work function was compared with results extracted from electronic structure calcu-lations of W(100) with commensurate Na overlayers at different coverage using the pseudopotential plane wave method based on density functional theory (DFT). Lang[7] carried out surface electronic structure calculations of a metal surface with a single Na atom, and Ishida[8] calculated the electronic structure of Na interfaces at different Na coverage, both applying the jellium model to the substrate. These results[7, 8] were used in Ref. [15] to interpret the experimental FE and PFE data. The low-index surfaces of W, however, contain several *d*-like surface states and surface resonances[16,17] that are crucial for the understanding of the substrate-overlayer bonding, and which the jellium model was not able to accurately describe.

In the present paper, we report the modifications in the surface electronic structures of W(100), W(110) and W(111) substrates due to commensurate Na overlayers using the *ab-initio* full-potential linear augmented plane wave (FP-LAPW) method in order to interpret the existing experimental FE and PFE data[15]. FE and PFE are very sensitive to the electronic states close to the center of the surface Brillouin zone (SBZ), because the surface potential barrier attenuates exponentially the electronic states of increasing wavevector. To accurately compare the theoretical results with the experiments we extend the electronic structure calculations to yield the total energy distributions of the emission currents (TEDs) in FE and surface PFE and the *k*-resolved layer densities of states (K-LDOS).

Low-energy electron diffraction (LEED) measurements[11] demonstrate that Na from an atomic beam grows pseudomorphically layer by layer on W(100)/Na up to 80 layers in a c(2x2) structure corresponding to $\theta = 1/2$. Our W(100) calculations were carried out with a Na c(2×2) overlayer [denoted W(100)/Na c(2×2)], which corresponds to one monolayer (ML) coverage. On W(110), LEED data[10] show that the Na overlayer undergoes a number of successive commensurate structures before completing the first atomic ML, which is an incommensurate, hexagonal layer at a coverage corresponding to $\theta = 3/5$. Since our numerical method, based on the repeated supercell geometry, involves the assumption of translational symmetry parallel to the surface, only commensurate overlayers were considered. We carried out emission current calculations at an commensurate overlayer corresponding to $\theta = 1/4$ (denoted W(110)/Na ($S_{1/4}$) , *S* stands for structure), which represents a (2×2) overlayer observed in the LEED data[10], in order to compare them with available experimental results[15]. To study the modification of the electronic structure with increased Na concentration we also report results of W(110) substrates with overlayers at $\theta = 2/5$ [W(110)/Na ($S_{2/5}$)], which is the commensurate overlayer of highest density observed by the LEED measurements, and at $\theta = 1/6$ [W(110)/Na ($S_{1/6}$)], which is the lowest-coverage commensurate overlayer observed. Regarding the Na overlayer on W(111), despite that W has a smaller atomic radius than Na, the low atomic density of the W(111) substrate allows it, in principle, to accommodate a Na (1x1) overlayer of equal atomic density. Our W(111) calculations were carried out with a hexagonal Na (1×1) overlayer [denoted W(111)/Na (1×1)]  To the best of our knowledge, no experimental data on the structure of Na overlayers on W(111) exists.

The electronic structure calculations are outlined in Section II. The method of calculation has been described in more detail elsewhere[16]. In Section III the calculations of W(100)/Na c(2×2) are interpreted and compared to the experimentally measured TEDs in FE and PFE. Similar discussions of W(110)/Na ($S_{2/5}$), W(110)/Na ($S_{1/4}$), W(110)/Na ($S_{1/6}$) and W(111)/Na (1×1) follow in Sections IV and V respectively. The results and conclusions of this work are summarized in Section VI.

## II. COMPUTATIONAL METHODS

The metal-adsorbate-vacuum interface is represented by a supercell, whose electronic structure is calculated self-consistently on the basis of density functional theory by the FP-LAPW method. Exchange and correlation are treated in the generalized-gradient approximation[18], and relativistic corrections including the spin-orbit interaction are taken into account unless otherwise specified. We used the electronic structure software package WIEN2K[19], which was modified as described in Refs. 16 and 17 to calculate the TEDs in FE and surface PFE, and the ratio of the TED of the metal to that of the free-electron metal (the enhancement factor of the emission current). We also report the ***k***-resolved surface density of states (K-SDOS), which describes the surface density of states (SDOS) weighted by the normalized tunneling factor, hence emphasizing the features of the electronic structure in the vicinity of the center of the SBZ ($\bar{\Gamma}$).

We found that for satisfactory convergence of our surface electronic structure calculations of clean W, the supercell must have at least two intermediate layers in register with the surface and the bulk. The geometry of the W(100)/Na c(2×2) interface is therefore described by a supercell that consists of 13 W layers stacked parallel to the (100) plane, surrounded on each side by a region of half that volume containing one Na atom. The W(110)/Na ($S_{1/4}$) supercell, corresponding to a (2×2) overlayer, consists of 13 W layers stacked parallel to the (110) plane, and the W(111)/Na (1×1) supercell consists of 19 W layers stacked parallel to the W(111) plane surrounded on each side by a region of half that volume containing one Na atom.

In Fig. 1a, 2a and 3a the solid square and rhombuses represent primitive unit cells of the Na c(2×2), (2×2) and (1×1) overlayers, and the dashed square and rhombuses represent primitive unit cells of the clean W(100), W(110) and W(111) substrates respectively. The corresponding SBZs are shown in Fig. 1b, 2b and 3b respectively. The symmetry points of the SBZ of the overlayers are denoted by a prime. The effect of the c(2×2) overlayer is to fold back the symmetry point $\bar{M}$ of the (1×1) substrate to $\bar{\Gamma}'$ in the SBZ of the overlayer. The effect of the (2×2) overlayer is to fold back the symmetry points $\bar{N}$ and $\bar{S}$ of the substrate to $\bar{\Gamma}'$ in the SBZ of the overlayer.

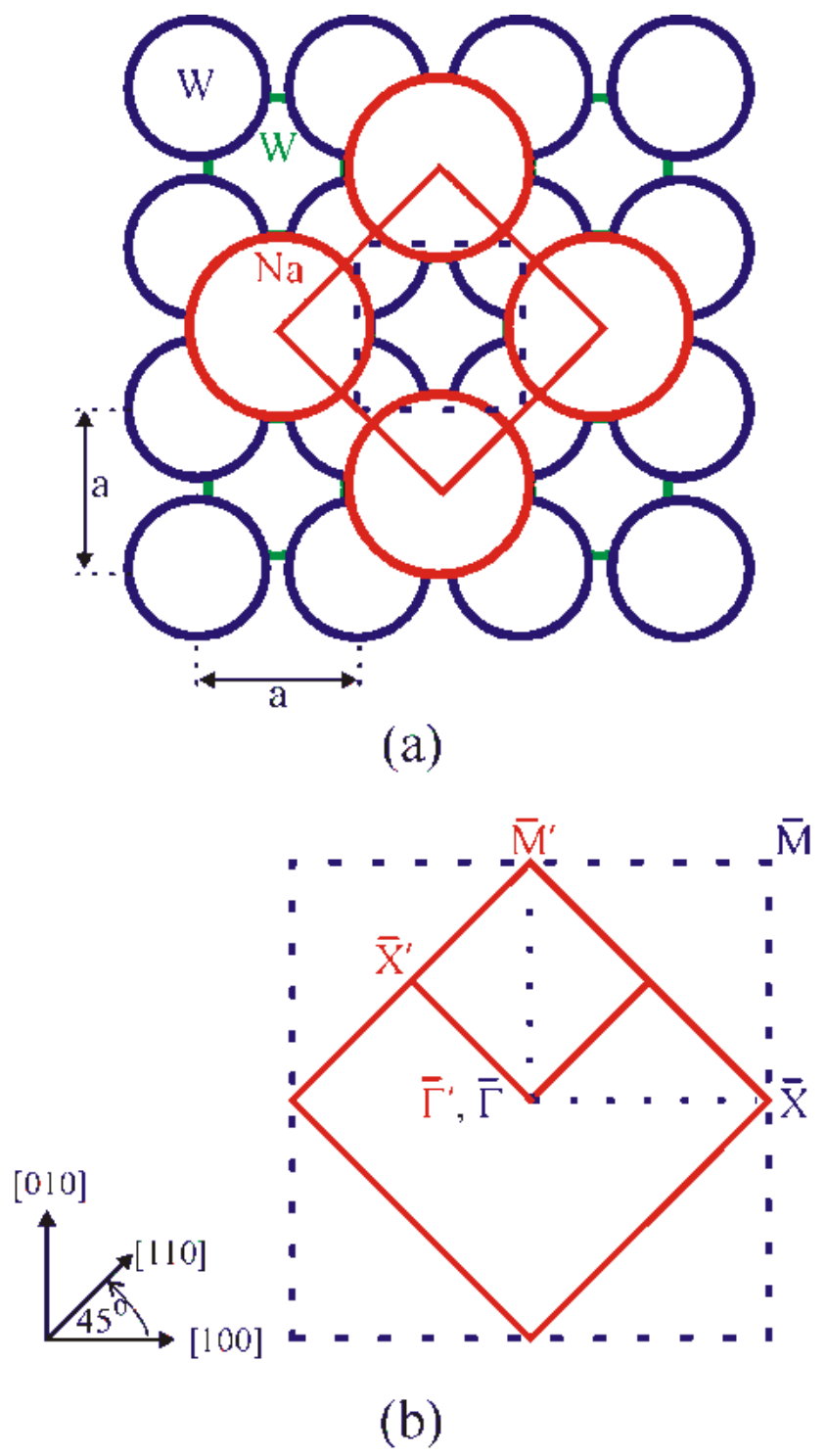

Fig. 1. (Color online) Surface unit cells and surface Brillouin zones of clean W(100) and of a Na c(2×2) overlayer on W(100). (a) Atomic positions in the top two layers of W(100) and in the overlayer. Also shown are the primitive unit cells of the clean substrate (blue dashed lines) and of the overlayer (red solid lines). (b) The surface Brillouin zones of the substrate and the overlayer plotted in the correct orientation relative to the unit cells in (a).

In interpreting the PFE results, it is important to distinguish between electron transitions resulting from *p* and *s* polarization of the incident light. In *p* polarization, electrons are excited by surface photoexcitation, so the final states are a continuum of free-electron-like states just outside the surface. Hence any features in the calculated TEDs in PFE in surface photoexcitation correspond to the initial states of the optical transition. If the polarized beam is also partially *s* polarized, electrons are also excited by bulk photoexcitation, hence direct transitions are induced between the initial and final states inside the metal. Hence any features observed in bulk photoexcitation correspond to the initial and final states of the optical transition. Because the available experimental PFE data are carried out predominantly in *p* polarization and because the transition rate in surface photoexcitation is typically much stronger than that in bulk photoexcitation we limited our PFE calculations to surface photoexcitation. This might underestimate the strengths of some peaks in the TEDs in PFE that are predominantly due to bulk photoexcitation.

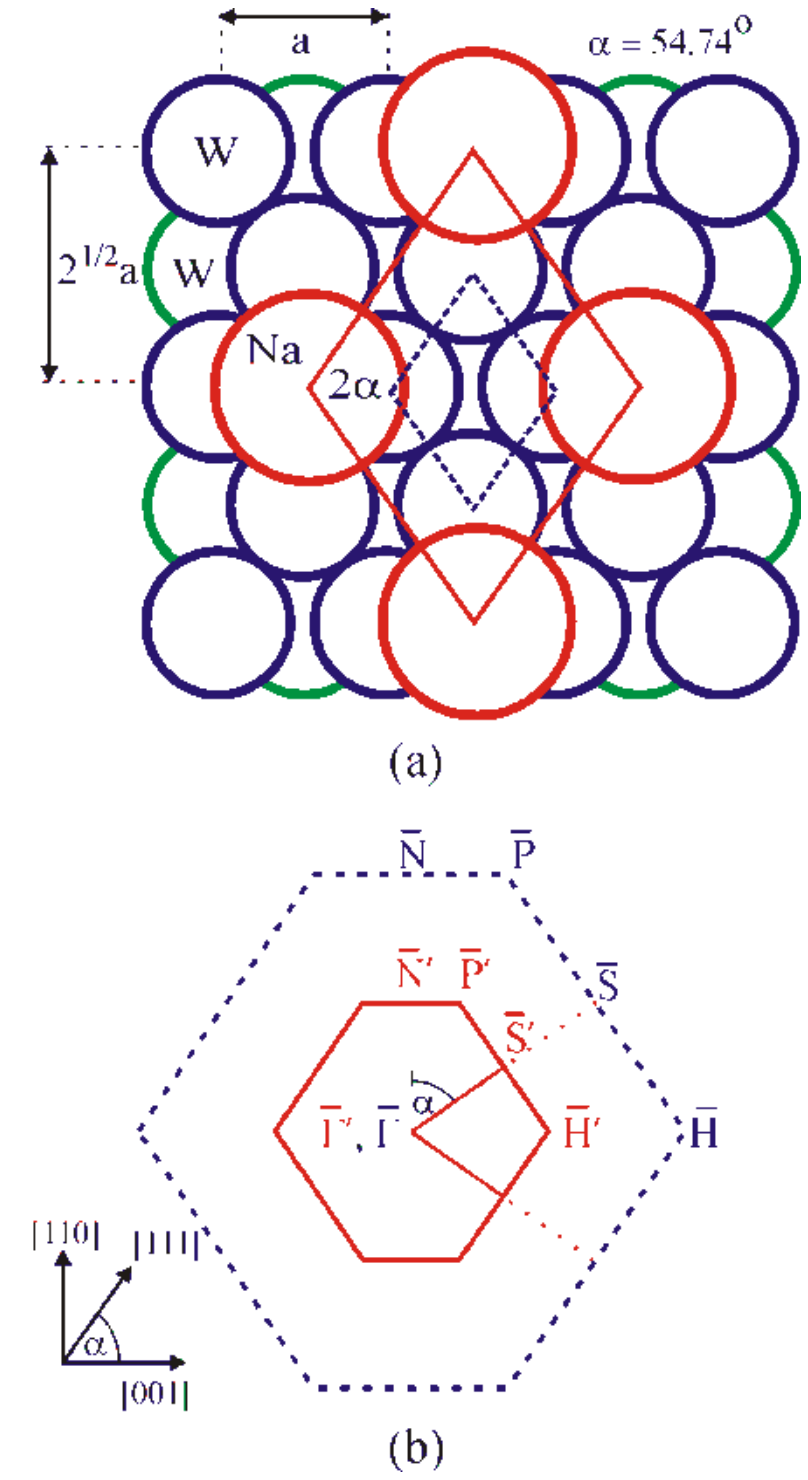

Fig. 2. (Color online) Surface unit cells and surface Brillouin zones of clean W(110) and of a Na ($S_{1/4}$) overlayer on W(110). Color and line style descriptions are the same as in Fig. 1.

In this paper, the experimentally-observed emission peaks are labeled alphabetically and the emitting facet is denoted by a numerical subscript (1 denotes 100, 2 denotes 110 and 3 denotes 111). The calculated peaks are denoted by a prime. All energies are expressed relative to the Fermi level $E_F$.

## III. RESULTS AND DISCUSSION FOR W(100)/Na

### III.1. Field and photofield emission currents from W(100)/Na c(2×2)

Experimental TEDs in FE and PFE from clean W(100) at 70 K[20] and at room temperature[15,16,17,21] show a strong peak $B_1$, known as the Swanson hump[22]. The calculated dispersion plot of clean W(100) plotted in the SBZ of a c(2×2) structure is shown in Fig. 4a. The calculated K-SDOS of clean W(100), plotted in Fig. 5a, shows a peak $B_1'$ that is due to a pair of surface states $B_1'$ of $d_z{}^2$-like symmetry close to $\bar{\Gamma}'$ (Fig. 4a) that are responsible for the observed strong peak $B_1$, where z denotes the direction along the normal to the surface. The energies of the calculated and experimental peaks are compared in Table I, and the symmetries of the calculated electron states that dominate the emission are reported. A detailed analysis of FE and PFE from clean W(100) can be found in Ref. 16.

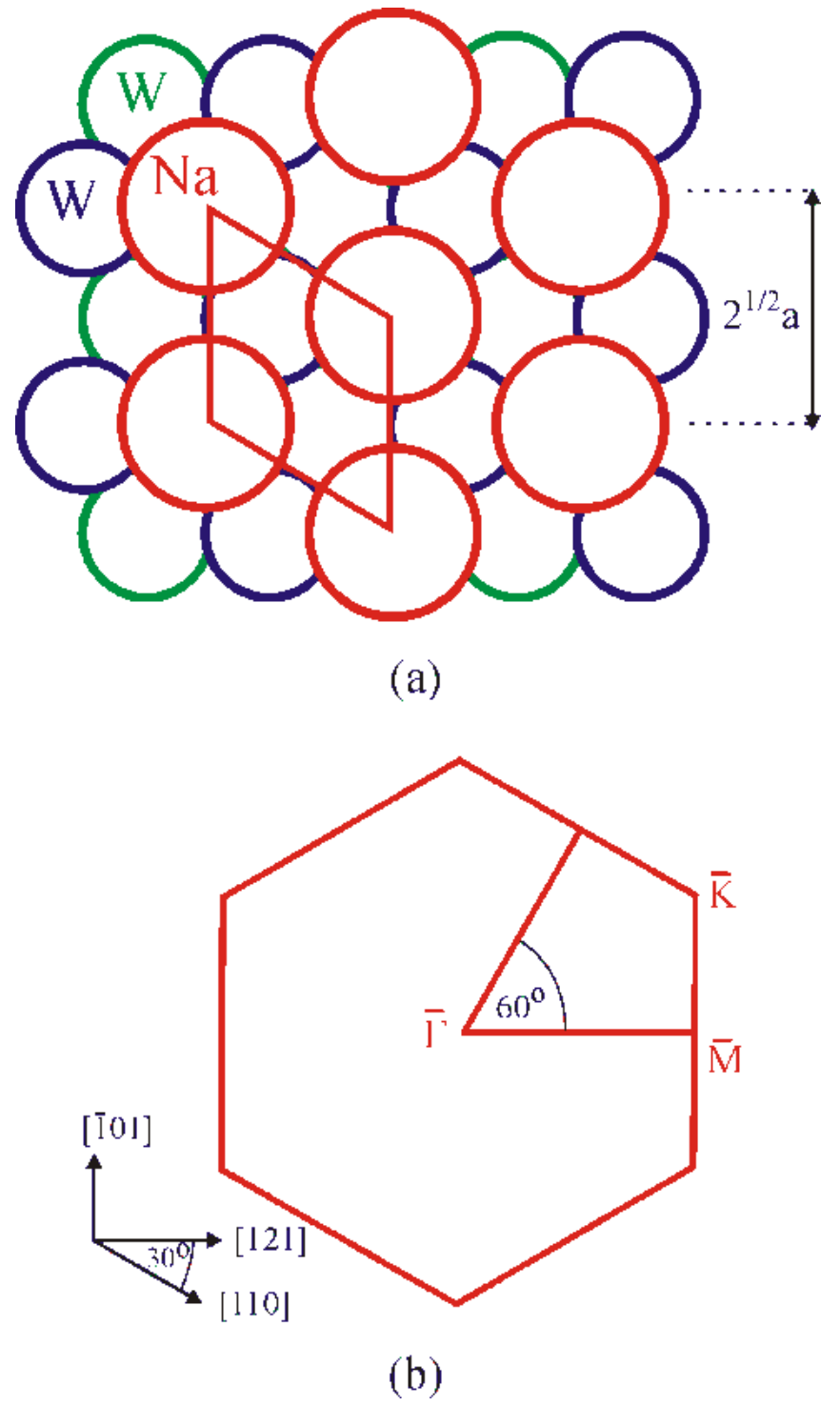

Fig. 3. (Color online) Surface unit cell and surface Brillouin zone of clean W(111) and of a Na (1×1) overlayer on W(111). (a) Atomic positions in the top two hexagonal layers of W(111) and in the overlayer. Also shown is the primitive surface unit cell of the clean substrate and of the overlayer. (b) The surface Brillouin zone plotted in correct orientation with respect to the corresponding unit cell in (a).

The logarithms of the calculated and the experimental[15] TEDs of the PFE current from W(100)/Na c(2×2) at room temperature as a function of the initial state energy are shown in Figs. 5c and 5d (upper curve at 1 ML coverage) respectively. In our calculations the energy of the photons in a *p*-polarized beam was 3.05 eV and the electric field strength was 0.16 V.Å$^{-1}$ as determined from the experimental data. The energies and symmetries of the peaks in FE and PFE are reported in Table 1. The band structure of W(100)/Na c(2×2) along $\bar{\Gamma}'\bar{X}'$ are shown in Fig. 4c and the band structure of an isolated Na layer (Fig. 4b) calculated using a supercell in which Na atoms occupy the same sites as in the W(100)/Na c(2×2) supercell and the W sites are empty is shown in Fig. 4b. When a Na c(2×2) overlayer is adsorbed on the clean W(100) substrate the *s*-like states of the overlayer near $\bar{\Gamma}'$ (Fig. 4b) hybridize with the surface states $B_1'$ of the substrate at –0.27 eV and –0.30 eV (Fig. 4a), shifting them to –1.49 eV (Fig. 4c) and yielding peaks $F_1'$ in the K-SDOS (Fig. 5b) and in the TED in PFE (Fig. 5c). This is consistent with the observed suppression of the experimental peak $B_1$ in the TED in FE from clean W(100) above a Na coverage of 0.5 ML, and the strong enhanced emission in the TED in FE at 1 ML at –1.5 eV, the lowest sampled energy, confirming a strong peak $F_1$ at or below –1.5 eV[15]. In Fig. 5d, the experimental TED in PFE at 1 ML shows a strong peak $F_1$ that is partially attributed to peak $F_1'$ calculated in surface photoexcitation.

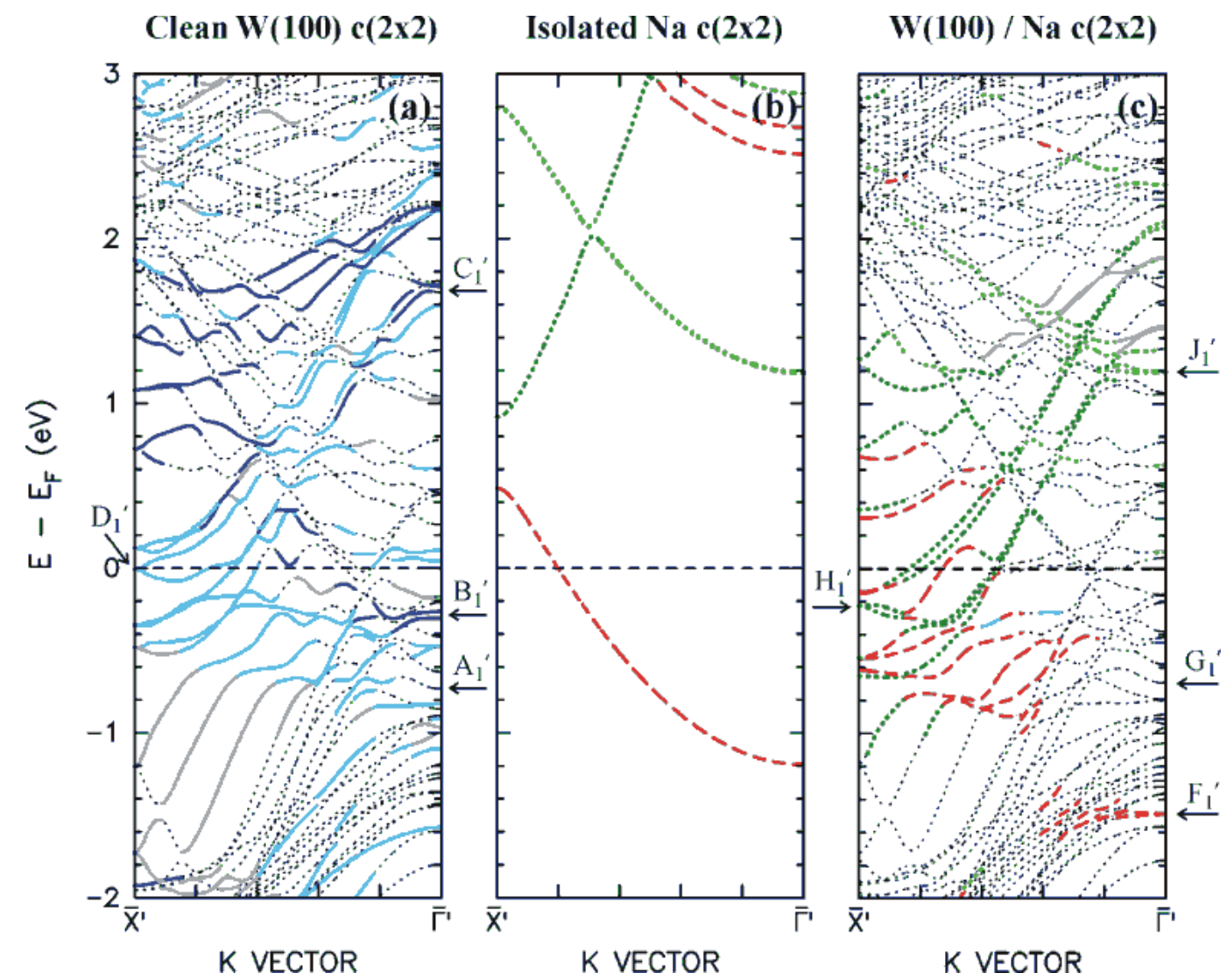

Fig. 4. (Color online) Dispersion along $\bar{X}'\bar{\Gamma}'$ of (a) clean W(100), (b) an isolated Na c(2×2) overlayer, and (c) W(100)/Na c(2×2), all plotted in the SBZ of the overlayer. Surface states and surface resonances are shown by thick curves. The predominant symmetry in the surface layer is shown by the line style [*s*, red (dashed); *p*, green (dotted); and *d*, blue or grey (solid)]. Bulk and intermediate states are shown by thin grey (dotted) curves.

In Fig. 4a the bands of clean W(100) are plotted in the SBZ of the c(2×2) overlayer in order to visualize the energy shifts when comparing them with the bands of W(100)/Na c(2×2). As a result the symmetry point $\bar{M}$ in the SBZ of clean W(100) is folded back to $\bar{\Gamma}'$ in the SBZ of the overlayer. This effect is responsible for the $d_{z^2}$-like surface states $C_1'$ with energies +1.7 eV and +2.18 eV at $\bar{\Gamma}'$ in Fig. 4a. Because FE is dominated by emission from electron states close to $\bar{\Gamma}'$, these states at $\bar{M}$ of the (1×1) W(100) substrate are expected to make a negligible contribution to emission from clean W(100) but can make a significant contribution to emission from W(100)/Na c(2×2) in bulk photoexcitation. In W(100)/Na c(2×2), the $p_z$-like states of the overlayer hybridize with the surface states $C_1'$ shifting them to lower energy to yield the strong peaks $J_1'$ in the K-SDOS. The experimental enhancement factor in FE at 1 ML coverage shows a strong peak $J_1$ at or above +1.0 eV[15] that is attributed to the surface states $J_1'$. The exact energy of peak $J_1$ cannot be determined because the exponential cutoff in the TED in FE that is due to the reduced electron occupation above $E_F$ prevents analyzing the data above +1.0 eV. Because the energies of states $F_1'$ and $J_1'$ differ by roughly the photon energy used, it is expected that bulk photoexcitation between these states will contribute to the strength of the experimental peak $F_1$. The strong peak $F_1$ observed in PFE is therefore attributed to surface photoexcitation from the initial states $F_1'$, as well as bulk pho-toexcitation between the initial states $F_1'$ and final states $J_1'$. Peak $A_1'$ in the calculated K-SDOS of clean W(100), due to a region of $d_{xz}+d_{yz}$-like surface resonances $A_1'$ at about −0.7 eV close to $\bar{\Gamma}'$, is responsible for the weak peak $A_1$ observed in the experimental TED in FE[16]. While peak $A_1'$ and $B_1'$ are comparable in the K-SDOS, peak $A_1'$ is much weaker in the TED in FE than peak $B_1'$, consistent with experiments, due to the different symmetry of the underlying electron states. In FE only the term of smallest parallel wavevector ($\mathbf{k}_{\|}+\mathbf{G}_{\|}$) in the plane wave expansion of the wave function contributes effectively to the emission current; $\mathbf{k}_{\|}$ denotes a wavevector in the first SBZ and $\mathbf{G}_{\|}$ denotes a reciprocal lattice vector[16,17]. While this term is significant in $s$-, $p_z$-, and $d_z^2$-like states at $\bar{\Gamma}$, because the wave functions of these states are symmetric in a plane parallel to the surface, it is small or zero in $p_x+p_y$-, $d_{xz}+d_{yz}$-, $d_{x^2-y^2}$- and $d_{xy}$-like states, because the wave functions are antisymmetric in a plane parallel to the surface. The $s$-like states of the overlayer hybridize with the $d_{xz}+d_{yz}$-like surface resonances $A_1'$ in an extended region along $\bar{\Gamma}'\bar{X}'$ to yield peak $G_1'$ in the calculated TED in PFE. The experimental TED in PFE at a Na coverage from 0.4 to 0.8 ML[15] show a strong peak $G_1$ (see lower curve in Fig. 5d at 0.6 ML coverage) that is attributed to the electron states $G_1'$.

Ref. 15 also suggests a peak labeled D at a final state energy at +3.2 eV in the TED in PFE that was attributed to a high DOS at the final state energy. In Fig. 5 of Ref. 15, however, the TED in PFE with 2.61 eV photons at 1 ML was shifted erroneously in energy by +0.5 eV resulting in the suggestion that peak D was due to a high DOS at the final state energy, i.e. due to bulk photoexcitation. Our calculations show that at $\bar{X}'$, the $p_x+p_y$ -like states of the overlayer hybridize with the $d_z^2$ –like surface states $D_1'$ of clean W(100) above $E_F$ shifting them to –0.22 eV. The extensions of these bands of surface states and surface resonances that disperse along $\bar{X}'\bar{\Gamma}'$ to higher energy crossing $E_F$ yield peak $H_1'$. The weak peak $H_1$ (labeled D in Ref. 15) in the experimental TED in PFE at 1 ML Na coverage is attributed to peak $H_1'$ calculated in surface photoexcitation.

### III.2. Layer densities of states of W(100)/Na c(2×2)

The layer densities of states (LDOSs) in the central (bulk) layer and in the surface layer (SDOS) of the supercell of clean W(100), shown in Figs. 6a and 6b respectively, show large differences due to the breaking of translational symmetry and the reduced atomic coordination at the surface. Strong peaks in the bulk layer are suppressed at the surface, and strong surface resonance peaks at about –4.1 eV, –0.3 eV and +1.6 eV are suppressed in the bulk.

The LDOS of an isolated Na layer, calculated using a supercell in which Na atoms occupy the same sites as in the W(100)/Na c(2×2) supercell and the W sites are empty, is shown in Fig. 6c. The LDOS is nearly free-electron-like below $E_F$. From –1.2 eV to +1.0 eV it is predominantly $s$-like with minor admixture of $p_x+p_y$-like states and above +1.0 eV it is predominantly $p$-like. The LDOSs in the substrate and in the overlayer of the supercell of W(100)/Na c(2×2) are shown in Figs. 6d and 6e respectively. Due to hybridization of the $s$- and $p$-like states of the overlayer with the states of the substrate the LDOS in the overlayer deviates strongly from its nearly free-electron-like nature in the isolated state. It spreads out to −4.5 eV and shows strong $s$-like peaks at –4.1eV and –1.5 eV, and $p_x+p_y$-like peaks at –0.3 eV and +1.1 eV due to the interaction of the Na states with the W surface resonances. The strong surface resonance peaks of W(100) below $E_F$ are only slightly modified thereby.

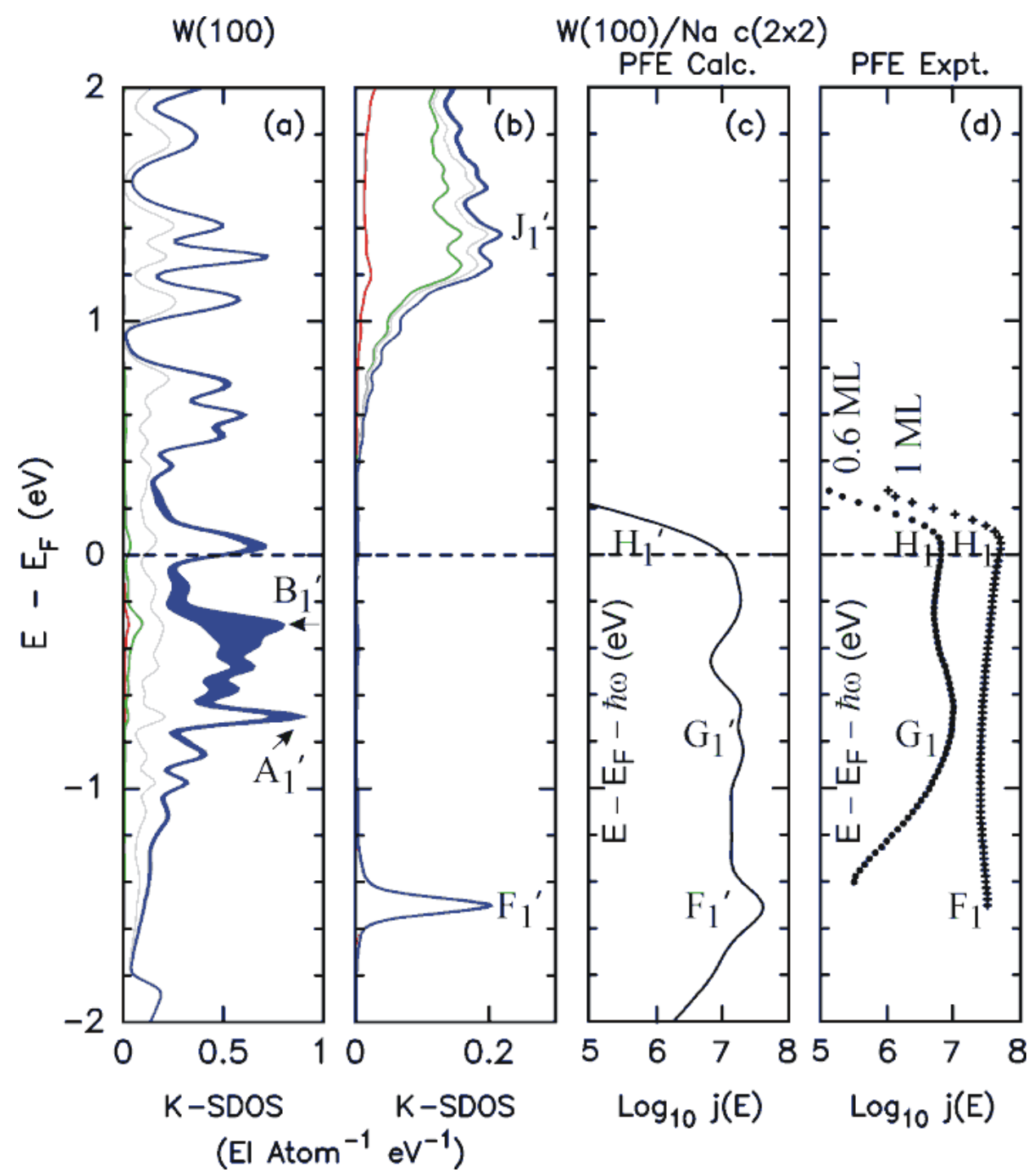

Fig. 5. (Color online) (a) K-SDOS of clean W(100) and (b) K-SDOS of a Na c(2×2) overlayer on W(100). The successive curves in the cumulative plots (a) and (b) show the contributions of $s$-like (red), $p$-like (green), $d_{xy}+d_{x^2-y^2}$-like (grey), $d_{xz}+d_{yz}$-like (light blue) and $d_z^2$-like (dark blue) states, respectively. The shading denotes the $d_z^2$-like contributions. [(c) and (d)] TEDs in PFE for W(100)/Na c(2×2) with 3.05 eV photons, plotted as a function of the initial state energy. The calculated plot (c) is based on surface photoexcitation. The experimental plot (d) shows a strong peak $F_1$ that is due to surface as well as bulk photoexcitation.

In summary, our calculated results are consistent with the peaks observed in FE and PFE from clean W(100) and W(100)/Na c(2×2) (Table I), indicating that the present calculations give a realistic picture of the changes in the electronic structure of W(100) due to adsorption of a Na c(2x2) overlayer.

For the muffin-tin radius of 1.7 Å used in the present calculation the total valence charge of 1.00 electrons within the muffin-tin sphere of Na does not change significantly upon adsorption. Instead, the total charge is redistributed among the various symmetry components. The *s*-like charge (integrated LDOS) decreases by about 0.43 electrons per Na atom and the *p*-and *d*-like charge increase by about 0.31 and 0.12 electrons per Na atom respectively. Hence, nearly no charge transfer takes place from the overlayer to the substrate. The same conclusion was reported[13] based on experimental photoemission measurements from W(110) with Na, K and Cs overlayers at coverages up to one atomic layer that show little if any shifts in binding energies (energy with respect to the Fermi energy) of the core 3p and 4f W-electrons.

Table I. Comparison of the peaks observed in the experimental TEDs in FE and PFE from clean W(100) ($A_1$ and $B_1$)[16,17] and W(100)/Na c(2×2) ($F_1$ to $J_1$)[15], with peaks in the calculated TEDs. *s.s.* stands for surface states and *s.r.* for surface resonances.

| Experimental | | Calculated | | | |
|---|---|---|---|---|---|
| Peak Label | $E–E_F$ (eV) | Peak Label | $E–E_F$ (eV) | Character | Symmetry in Overlayer [Substrate] |
| $A_1$ | −0.73 (5) | $A_1'$ | −0.69 (2) | *s.r.* | $[d_{xz}+d_{yz}]$ |
| $B_1$ | −0.32 (3) | $B_1'$ | −0.32 (2) | *s.s.* | $[d_z^2]$ |
| $F_1$ | ≤ −1.5 | $F_1'$ | −1.50 (2) | *s.s.* | $s\ [d_z^2]$ |
| $G_1$ | −0.8 (1) | $G_1'$ | −0.63, −0.83 (2) | *s.r.* | $s\ [d_{xz}+d_{yz}]$ |
| $H_1$ | Close to +0.0 | $H_1'$ | −0.05+0.04 (2) | *s.r.* | $p_x+p_y\ [d_{xz}+d_{yz}]$ |
| $J_1$ | ≤+1.55 | $J_1'$ | +1.2 to +1.6 | *s.s. - s.r.* | $p_z\ [d_z^2]$ |

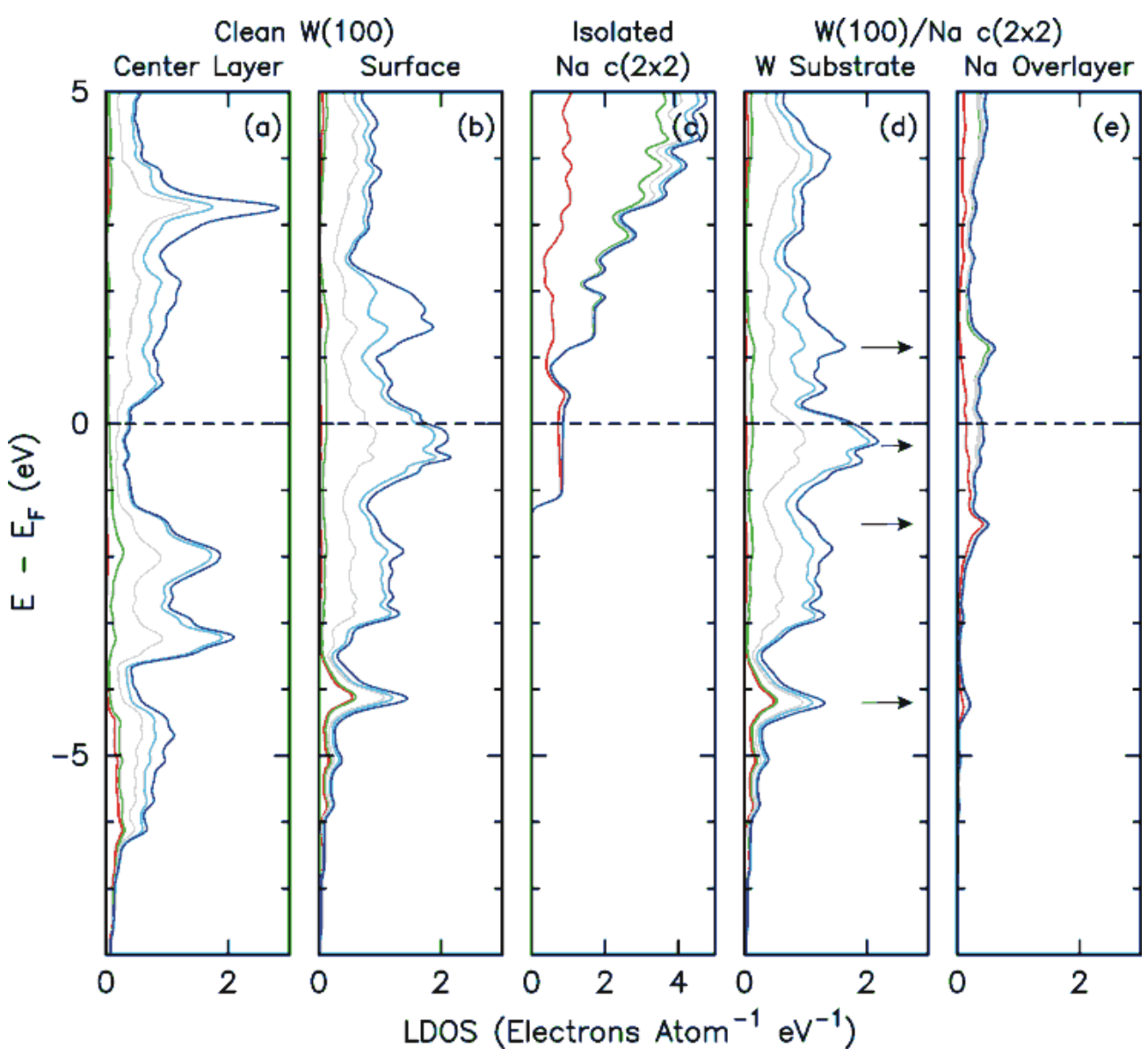

Fig. 6. (Color online) LDOS of clean W(100) in the (a) central (bulk) layer and (b) surface layer. (c) LDOS in an isolated Na c(2×2) layer. LDOS of W(100)/Na c(2×2) in the (d) W substrate and (e) Na overlayer. In these cumulative plots, the areas between successive curves show the contributions of *s*-like (red), *p*-like (green), $d_{xy}+d_{x^2-y^2}$-like (grey), $d_{xz}+d_{yz}$-like (light blue) and $d_z^2$-like (dark blue) states, respectively.

### III.3. Work function of W(100)/Na c(2×2)

Typically, valence electrons spill out from the surface of a clean transition metal, producing an inwardly-directed surface electric dipole layer that increases the work function. Alkali and alkaline-earth overlayers on transition metal surfaces produce additional surface electric dipole layers whose net effect is to decrease the work function.

The spatial redistribution of the valence electron density that occurs when a Na c(2×2) overlayer is adsorbed on W(100) was calculated by subtracting the sum of the valence electron density distribution of the clean W substrate and of the isolated Na c(2×2) overlayer from the valence electron density distribution of W(100)/Na c(2×2). The redistribution of the electron density in a {110} plane that passes through the metal surface is shown in Fig. 7. Electrons move out from the regions surrounding the W and Na atoms and accumulate in a layer between the overlayer and the substrate. The net effect is an outwardly-directed dipole layer resulting in a Na-induced lowering in the potential energy of an electron outside the surface with respect to the bulk and thus lowering the work function. In the present work, the work function of W(100)/Na c(2×2) is 2.1 eV as estimated by calculating the difference between the Coulomb potential energy far into the vacuum region of the supercell and the Fermi energy. The calculated work function agrees well with the value of 2.2 eV observed experimentally at 77 K using a field-emission microscope[9], and with the value of 2.3 eV calculated by means of the pseudopotential plane wave method based on density functional theory[3].

† Deceased 21 April 2009

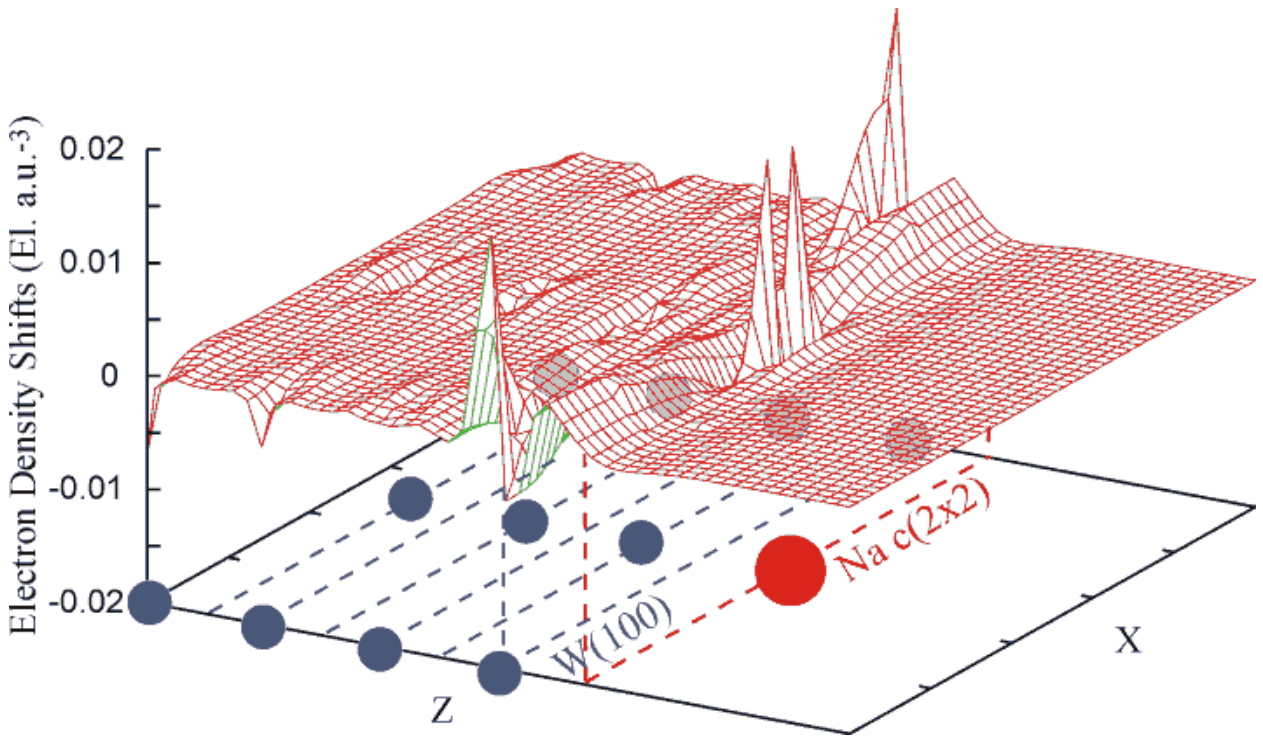

Fig. 7. (Color online) Calculated redistribution of the valence electron density that occurs when a Na c(2×2) overlayer is adsorbed on a W(100) surface, plotted in a {110} plane that intersects the surface at right angles. The dashed rectangles represent the planes of the overlayer (labeled Na) and of the substrate (labeled W). A positive shift corresponds to an increased electron density.

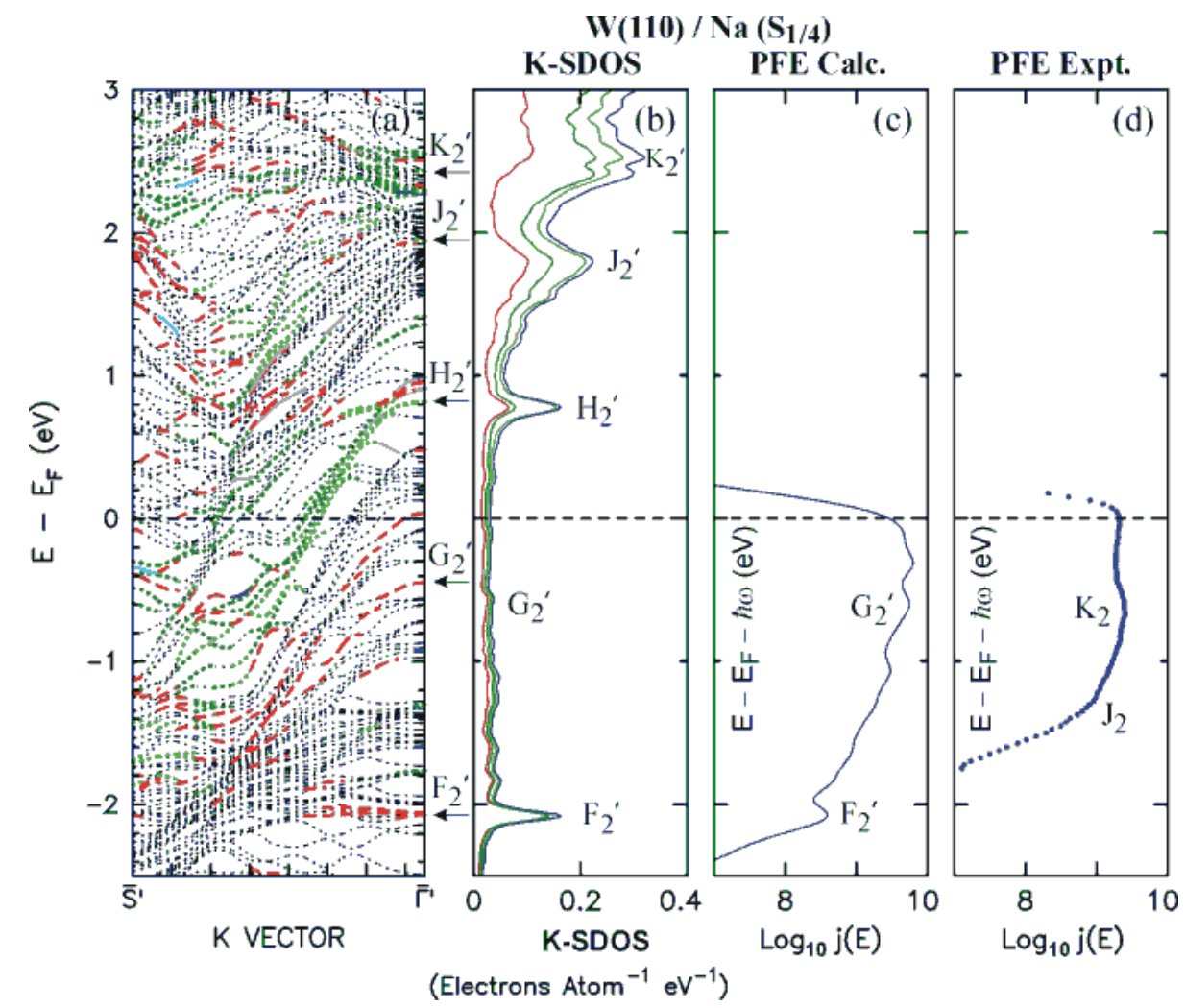

Fig. 8. (Color online) (a) Dispersion plots along $\bar{S}'\bar{\Gamma}'$ in the surface Brillouin zone of W(110)/Na ($S_{1/4}$). Surface states and surface resonances are shown by thick curves. The predominant symmetry in the surface layer is shown by the line style [*s*, red (dashed); *p*, green (dotted); and *d*, blue (solid)]. Bulk and intermediate states are shown by thin grey dotted curves. (b) K-SDOS of W(110)/Na ($S_{1/4}$). The successive curves in the cumulative plots show the contributions of *s*-like (red), $p_x$+$p_y$-like (dark green), $p_z$-like (light green) and *d*-like (blue) states, respectively. [(c) and (d)] TEDs in PFE for W(110)/Na ($S_{1/4}$) with 3.05 eV photons, plotted as a function of the initial state energy. The calculated plot (c) is based on surface photoexcitation. The experimental plot (d) shows two additional final state peaks $K_2$ and $J_2$ that are attributed to bulk photoexcitation.

## IV. RESULTS AND DISCUSSION FOR W(110)/Na

### IV.1. Field and photofield emission currents from W(110)/Na ($S_{1/4}$)

The calculated TEDs in FE and PFE from clean W(110)[17] are consistent with the experimental TEDs at 78K[20] and at room temperature[15,17] in that they show little initial state structure. The observed TEDs in FE in the range of Na coverage from $\theta = 0$ to 1/4 show no structure, but above that coverage they show several peaks[15]. In Table II, the energies and symmetries of the calculated peaks in the TEDs in PFE at 3.05 eV photons (Fig. 8c) and the K-SDOS of W(110)/Na ($S_{1/4}$) (Fig. 8b) are listed, and the energies of the peaks are compared with those observed in the PFE experiments (Fig. 8d).

To facilitate the interpretation of the charge shifts in W(110) due to the Na ($S_{1/4}$) overlayer the energy bands and LDOS (Fig. 9c) of an isolated Na layer calculated using a supercell in which Na atoms occupy the same sites as in the W(110)/Na ($S_{1/4}$) supercell and the W sites are empty have been calculated. The *s*-like valence states form a nearly-free-electron band along $\bar{\Gamma}'\bar{S}'$. The bottom of the band is at −0.79 eV. The $p_z$-like states of the isolated Na ($S_{1/4}$) layer form another band at +1.39 eV that disperses to higher energy along $\bar{\Gamma}'\bar{S}'$ while the $p_x$+$p_y$-like states form a band at +1.06 that disperses to higher energy along $\bar{S}'\bar{\Gamma}'$.

The TEDs in PFE with 3.05 eV photons in the range of coverage from $\theta = 1/7$ to $\theta = 1/4$ (Fig. 8d) show a strong peak $K_2$ that shifts slightly to lower energy with increased coverage, and that has been attributed to bulk photoexcitation[15]. At the corresponding final state energy of +2.45 eV the K-SDOS of W(110)/Na ($S_{1/4}$) shown in Fig. 8b shows a strong peak $K_2'$. In the presence of a (2×2) overlayer, the symmetry point $\bar{S}$ in the SBZ of clean W(110) is folded back to $\bar{\Gamma}'$ in the SBZ of the overlayer. Bands of $p_x$+$p_y$-like states of the isolated Na layer close to $\bar{\Gamma}'$ hybridize with $d_z{}^2$-like W states at +2.74 eV in clean W(110) that have been folded back from $\bar{S}$ to yield the bands of surface resonance states $K_2'$ (Fig. 8a). The calculated TED shows a weak peak $G_2'$ at the corresponding initial state energy that is due to a band of *s*-like surface resonances. The observed peak $K_2$ is attributed to bulk photoexcitation between these states. The observed TED in PFE shows another peak $J_2$ at 1 ML. Peak $J_2'$ in the K-SDOS is due to *s* and $p_z$-like states of the overlayer close to $\bar{\Gamma}'$ that hybridize with $d_z{}^2$-like states of the substrate at +1.89 eV. The observed peak $J_2$ is attributed to bulk photoexcitation to these states.

Table II. Comparison between the energies of the peaks observed in the TEDs in PFE from W(110)/Na ($S_{1/4}$)[15] and the energies of the calculated peaks. *s.r.* stands for surface resonances.

| Experimental | | Calculated | | | |
|---|---|---|---|---|---|
| Peak Label | $E$–$E_F$ (eV) | Peak Label | $E$–$E_F$ (eV) | Character | Symmetry in Overlayer [Substrate] |
| | | $F_2'$ | –2.08 (2) | *s.r.* | $s$ [$d_{xz}$+$d_{yz}$] |
| $G_2$ | –0.60 (3) | $G_2'$ | –0.60 (2) | *s.r.* | $s$ [$d_{xz}$+$d_{yz}$] |
| | | $H_2'$ | +0.80 (2) | *s.r.* | $p_z$ [$p_z$] |
| $J_2$ | +1.65 (3) | $J_2'$ | +1.80 (5) | *s.r.* | $s$, $p_z$ [$d_{z^2}$] |
| $K_2$ | +2.45 (3) | $K_2'$ | +2.5 (1) | *s.r.* | $p_x$+$p_y$ [$d_{z^2}$] |

The $s$-like states of the Na ($S_{1/4}$) overlayer close to $\bar{\Gamma}'$ hybridize with the W states to form bands of $s$-like surface resonances $F_2'$ that yield a strong peak $F_2'$ in the calculated TED in PFE at –2.08 eV. Because the energy of the peak of the surface potential barrier is about +1.55 eV in the PFE experiments of Fig. 8d, only electrons having normal energy greater than –1.50 eV can pass above the peak of the barrier which accounts for the cutoff below about –1.5 eV. Peak $F_2'$ could not be verified by the experimental results because its energy range is below the cutoff energy in the experimental TED.

### IV.2. Layer densities of states of W(110)/Na

Clean W(110) shows prominent peaks in the bulk LDOS (Fig. 9a) that are greatly attenuated in the surface layer (Fig. 9b) and shows prominent surface peaks at –3.0 eV, –1.5 eV and +2.3 eV, that grow progressively weaker in the subsurface layers. The LDOS of an isolated Na layer, calculated using a supercell in which Na atoms occupy the same sites as in the W(110)/Na ($S_{1/4}$) supercell and the W sites are empty, is shown in Fig. 9c. The LDOS is $s$-like below +0.8 eV and is well separated from the $p$-like LDOS that predominates above +1.0 eV. In W(110)/Na ($S_{1/4}$) overlap between the $s$ and $p$-like Na states and the W states causes the energy gap in the overlayer above $E_F$ (Fig. 9e) to disappear. The LDOS in the overlayer spreads out in energy to about –3.5 eV and shows a strong $s$-like peak at –2.1 eV and several predominantly $p$-like peaks above $E_F$. The strong surface resonance peaks of clean W(110) shift slightly in energy (Fig. 9d) and extend into the overlayer.

Peak $K_2$ in the experimental TEDs in PFE from W(110)/Na shifts to lower energy with increased coverage[15]. Electronic structure calculations of a metal substrate represented by a jellium model with Na overlayers arranged in a square lattice at different coverages by Ishida[8], on the other hand, show no shift in the SDOS to lower energy of the valence states with increased Na coverage. To study the effect of Na coverage on the SDOS, we report the results of two more W(110) calculations at different coverages; W(110)/Na ($S_{1/6}$) at $\theta$ = 1/6 has the lowest-coverage commensurate overlayer observed by LEED data, and W(110)/Na ($S_{2/5}$) at $\theta$ = 2/5 has the highest-coverage commensurate overlayer observed[10]. The structures of the overlayers correspond to structures b and f in Fig.4 of Ref. 10 respectively. The same normal distances between the Na overlayers and the W substrate in the three calculations was used. Due to the computational challenges introduced by the large number of 81 inequivalent atoms in the supercell, spin-orbit interaction has not been included in the calculations at $\theta$ = 1/6. Spin-orbit interaction in clean W(110) is responsible for minor shifts in energy of some energy states (a maximum of 0.2 eV for states below $E_F$), and in the much lighter Na they show negligible effect. While inclusion of spin-orbit interaction in W(110)/Na ($S_{1/6}$) is expected to cause splitting of some states, the resulting energy shifts, as judged by the clean W(110) case, are expected to be smaller than the energy shifts caused by the coverage variations discussed in the present work, and hence they do not invalidate the discussion and conclusions in the coming paragraphs.

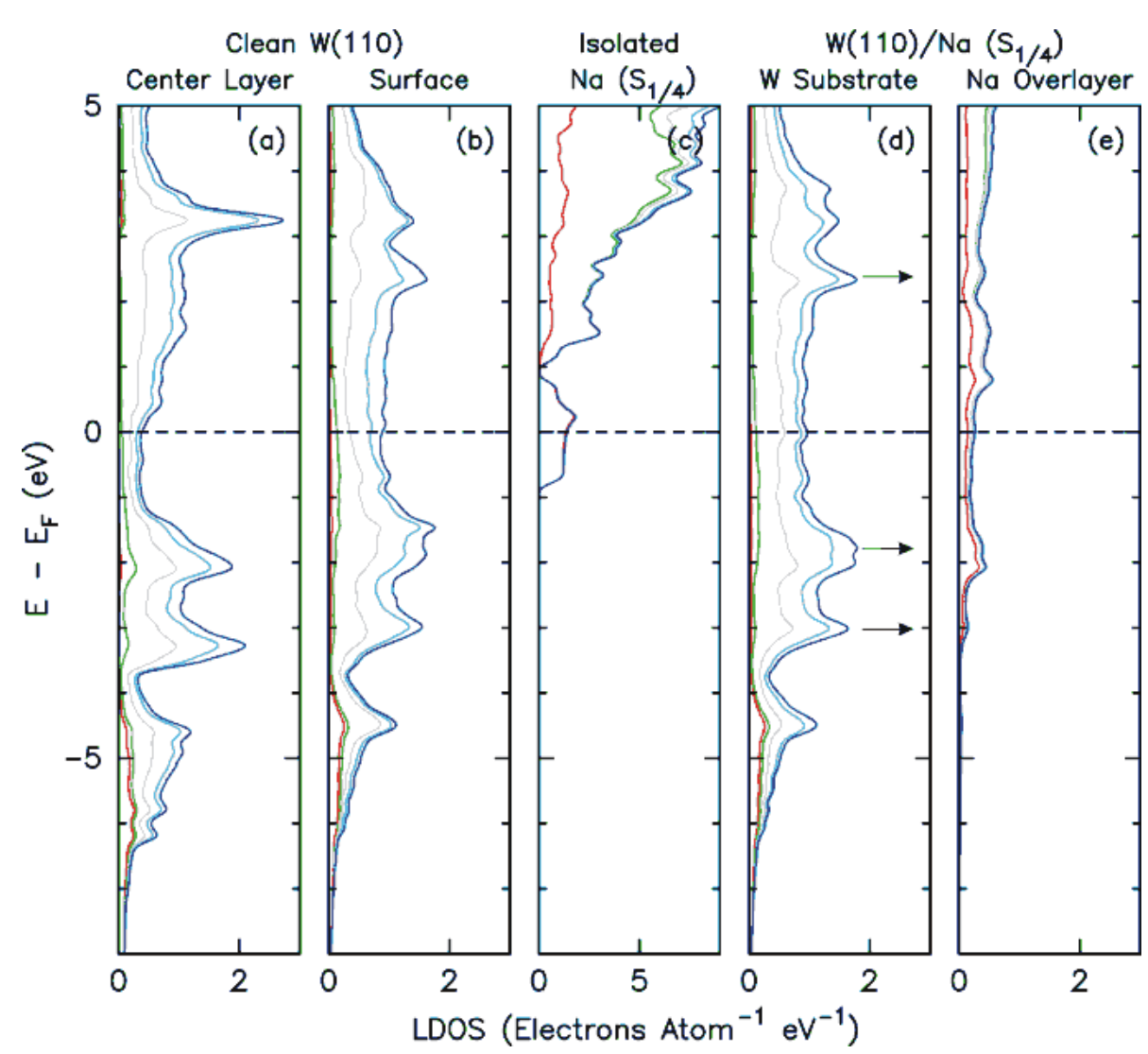

Fig. 9. (Color online) LDOS of clean W(110) in the (a) central (bulk) layer and (b) surface layer. (c) LDOS in an isolated Na ($S_{1/4}$) layer. LDOS of W(110)/Na ($S_{1/4}$) in the (d) W substrate and (e) Na overlayer. In these cumulative plots, the areas between successive curves show the contributions of $s$-like (red), $p$-like (green), $d_{xy}$+$d_{x^2-y^2}$-like (grey), $d_{xz}$+$d_{yz}$-like (light blue) and $d_{z^2}$-like (dark blue) states, respectively.

† Deceased 21 April 2009

The SDOSs in the three Na overlayers on W(110) of coverages $\theta = 1/6$, 1/4, 2/5, are shown in Figs. 10a, b and c. Overlap between the Na states and the W states results in a noticeable $p_x$, $p_y$ and $p_z$-like component in the SDOSs below $E_F$. Increasing the Na coverage broadens the peaks in the SDOS, especially above $E_F$. It also shifts the peaks to lower energy, for example, the *s*-like peak below $E_F$ at –2.0 eV shifts by about –0.4 eV when the coverage is increased from $\theta = 1/6$ to $\theta = 2/5$, and the strong *s* and $p_z$-like peak at +1.7 shifts by about –1.2 eV. Also the $p_x$+$p_y$-like peak at +2.5 eV shift by about –0.3 eV when the coverage is increased from $\theta = 1/4$ to $\theta = 2/5$, which is consistent with the experimental TEDs in PFE from W(110)/Na[15] that show a shift to lower energy of peak $K_2$ with increased coverage.

To separate the effect of intralayer Na-Na interactions from the interlayer W-Na interactions the LDOSs is reported for isolated Na layers using supercells in which Na atoms occupy the same sites as in the corresponding W(110)/Na supercells at coverages $\theta = 1/6$, 1/4, 2/5, and the W sites are empty (Figs. 11a, b and c respectively). At low Na concentration intralayer interactions between the 3*s* and 3*p* atomic electron states of Na on adjacent atoms broaden them into bands. This results in a *s*-like LDOS extending over a broad energy range through $E_F$ and a mainly *p*-like LDOS at higher energy. While in the isolated Na ($S_{1/6}$) and Na ($S_{1/4}$) layers the *s*-like states are well separated from the *p*-like states, in the isolated Na ($S_{2/5}$) layer overlap between the atomic *s*- and *p*-like states results in a continuous LDOS, and a weak $p_x$- and $p_y$-like component in the LDOS below $E_F$. It is seen that intralayer interactions shift, for example, the lowest energy *s*-like valence states in the LDOS of the isolated Na layers, which are states at $\bar{\Gamma}'$, by –1.0 eV when the Na concentration is increased from $\theta = 1/6$ to $\theta = 2/5$, and explains the decrease in energy of the strong *s*-like peak in the SDOS below $E_F$ with increasing coverage.

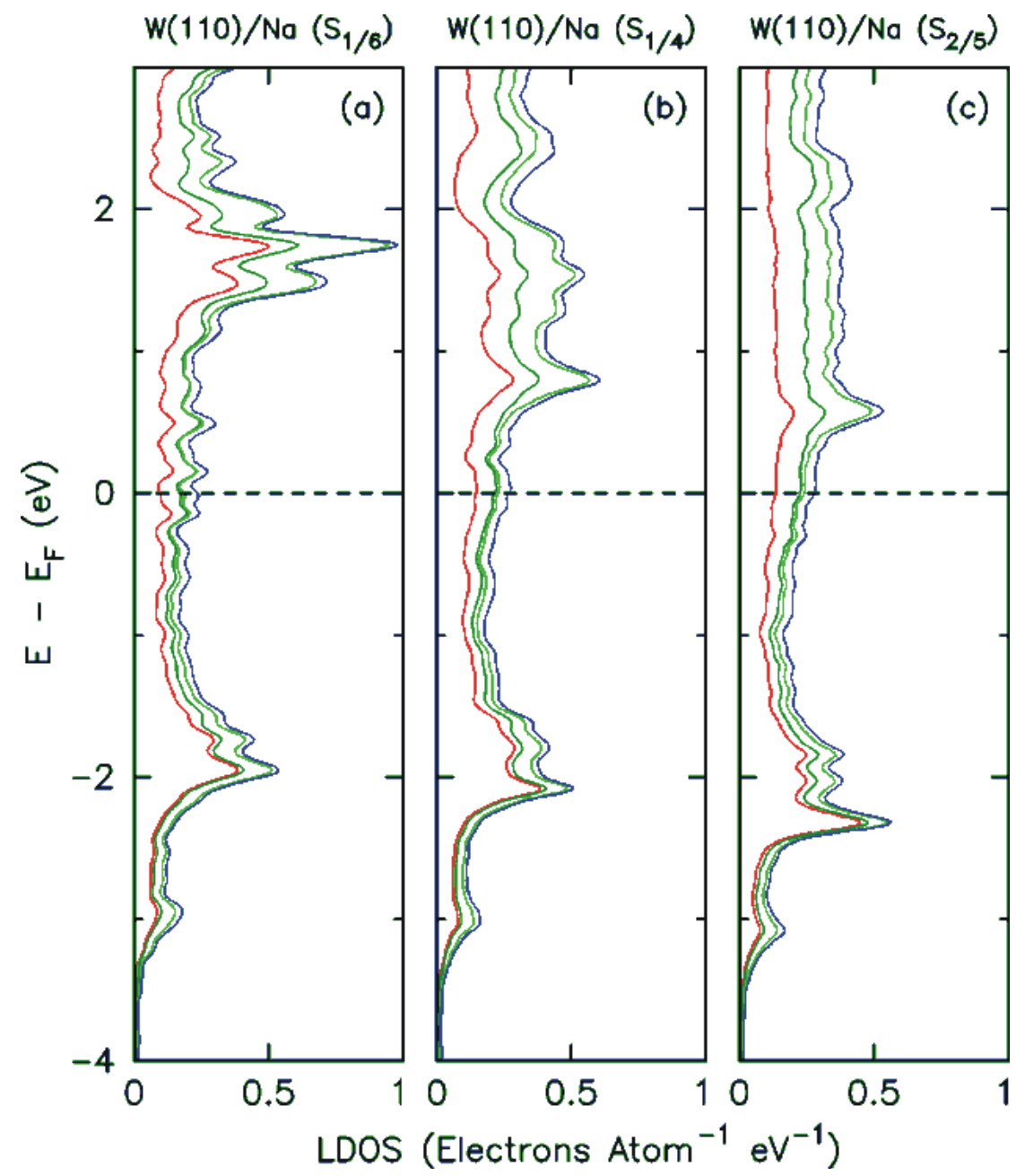

Fig. 10. (Color online) LDOS of W(110)/Na in the Na overlayer at increasing Na coverage for (a) W(110)/Na ($S_{1/6}$), (b) W(110)/Na ($S_{1/4}$) and (c) W(110)/Na ($S_{2/5}$). In these cumulative plots, the areas between successive curves show the contributions of *s*-like (red), $p_x$+$p_y$-like (dark green), $p_z$-like (light green) and *d*-like (blue) states, respectively.

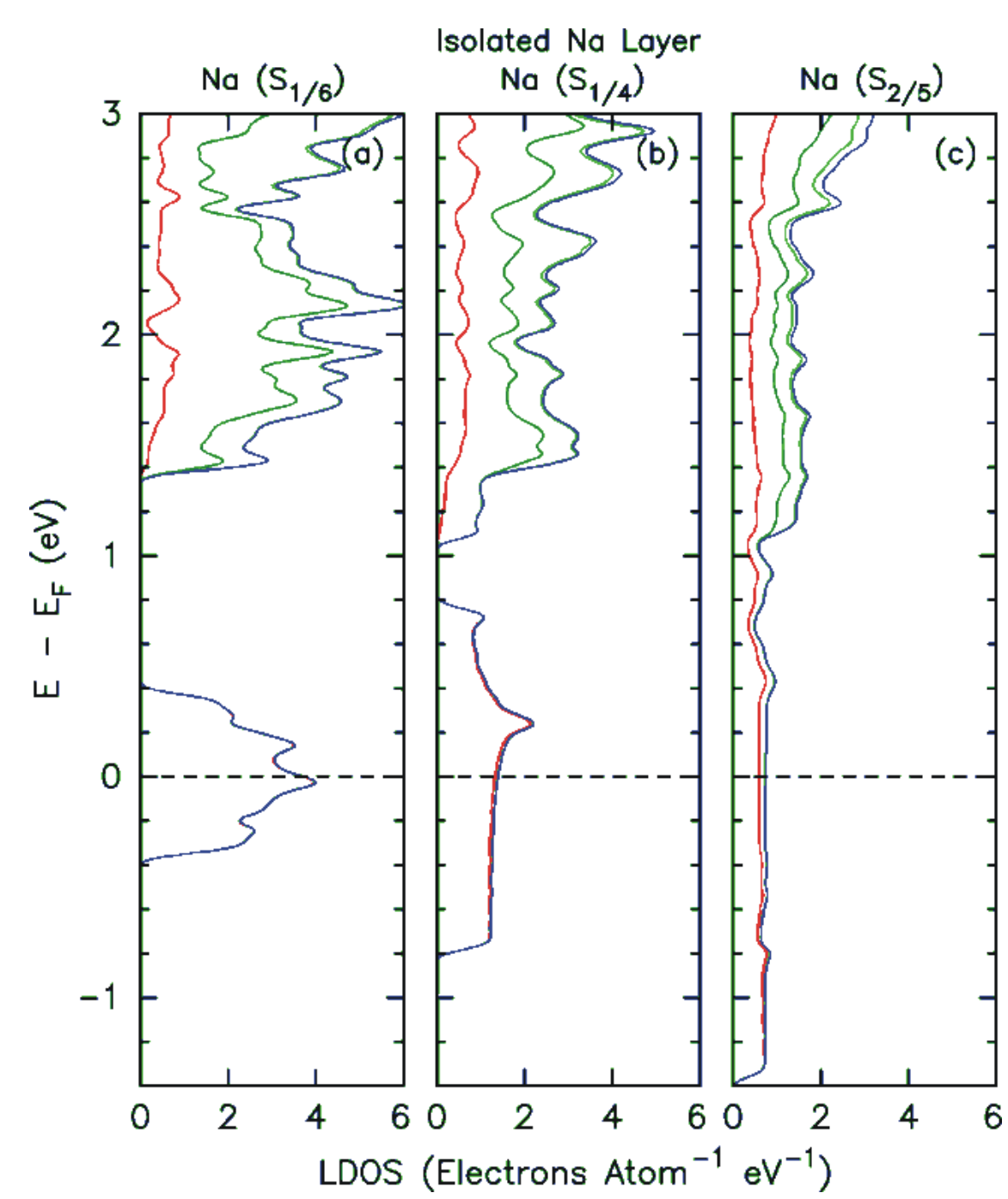

Fig. 11. (Color online) LDOS in isolated Na layers at increasing concentration for (a) Na ($S_{1/6}$), (b) Na ($S_{1/4}$) and (c) Na ($S_{2/5}$); the layer structures correspond to the overlayer structures in Fig. 10a, b and c, respectively. In these cumulative plots, the areas between successive curves show the contributions of *s*-like (red), $p_x$+$p_y$-like (dark green), $p_z$-like (light green) and *d*-like (blue) states, respectively.

## V. RESULTS AND DISCUSSION FOR W(111)/Na

### V.1. Field and photofield emission currents from W(111)/Na (1×1)

Measurements of the TEDs from clean W(111) at 78 K in FE[20] and at room temperature in FE and PFE[16,23] show a strong asymmetrical peak $B_3$ with maximum emission at about –0.7 eV, a broad peak $A_3$ centered at about –1.4 eV[20], and a rise in the experimental enhancement factor of clean W(111) just above $E_F$ that suggests the presence of a peak $C_3$. The calculated dispersion plot for clean W(111) (Fig. 12a) shows two bands of surface resonances of $d_z^2$-like symmetry that originate at $\bar{\Gamma}$ at about –0.57 eV and –0.42 eV respectively. The highly asymmetrical peak $B_3$ is attributed to emission from strongly lifetime broadened states $B_3'$ in these bands. Peak $C_3$ is consistent with the calculated asymmetrical peak $C_3'$ in FE that is due to a band of intermediate states of predominantly $d_z^2$-like symmetry that originates at $\bar{\Gamma}$ at +0.18 eV, and the broad peak $A_3$ is consistent with a calculated weak peak $A_3'$ in FE that is due to several bands of surface resonances close to $\bar{\Gamma}$. The energies and symmetries of the calculated peaks in the TEDs in FE and PFE from clean W(111) are listed in Table III and the energies are compared with those of the observed peaks. More details are given in Ref. 16.

The bottom of the band formed by the *s*-like valence states in an isolated Na layer, calculated using a supercell in which Na atoms occupy the same sites as in the W(111)/Na (1×1) supercell and the W sites are empty, is at –1.40 eV. It disperses to higher energy along $\bar{\Gamma}\bar{M}$ and crosses $E_F$ at $0.75\bar{\Gamma}\bar{M}$. At $\bar{M}$ the states in that band show $p_x$+$p_y$-like symmetry. At $\bar{\Gamma}$ the $p_z$-like conduction states form another band at +1.06 eV that disperses to higher energy along $\bar{\Gamma}\bar{M}$.

Fig. 12d shows the logarithm of the calculated TED of the PFE current from W(111)/Na (1×1) at room temperature, plotted as a function of the initial state energy. The photon energy was 3.05 eV, the electric field strength was 0.16 V.Å$^{-1}$, and the work function was 2.84 eV. The energy of the peak of the surface potential barrier is +1.32 eV. This accounts for the cutoff in the TED in PFE below about –1.7 eV. In Table III, the energies and symmetries of the calculated emission peaks in FE and PFE from W(111)/Na (1×1) are listed.

When the Na (1×1) overlayer is adsorbed on the clean W(111) substrate the *s*-like states of the overlayer hybridize with several W states near $\bar{\Gamma}$ resulting in a number of peaks in the K-SDOS of W(111)/Na ( 1×1) shown in Fig. 12c. The $d_z^2$-like surface resonances $B_3'$ of the clean substrate at –0.57 eV and –0.42 eV (Fig. 12a) shift to $F_3'$ at −1.48 eV and $G_3'$ at −0.98 eV (Fig. 12b) yielding peaks $F_3'$ and $G_3'$ in the K-SDOS. As a result the calculated peaks $B_3'$ of clean W(111) that yield the highly-asymmetrical peak $B_3$ observed in FE and PFE from clean W(111) are suppressed. The *d*-like surface resonances $A_3'$ and intermediate states of clean W(111) hybridize with the *s*-like and *p*-like states of the overlayer and yield several other peaks in the K-SDOS in the region from −1.70 eV to −0.98 eV. The TED in PFE in Fig. 12d shows a broad region of enhanced electron emission over that energy region due to emission from all these states. The unoccupied $d_z^2$-like states $C_3'$ of clean W(111) just above $E_F$ shift to below $E_F$ (labeled $H_3'$ in Fig. 12b) and yield peak $H_3'$ in the K-SDOS and in the TED in PFE. Unfortunately, there are no experimental data available in the literature on W(111)/Na to compare with our calculated TEDs.

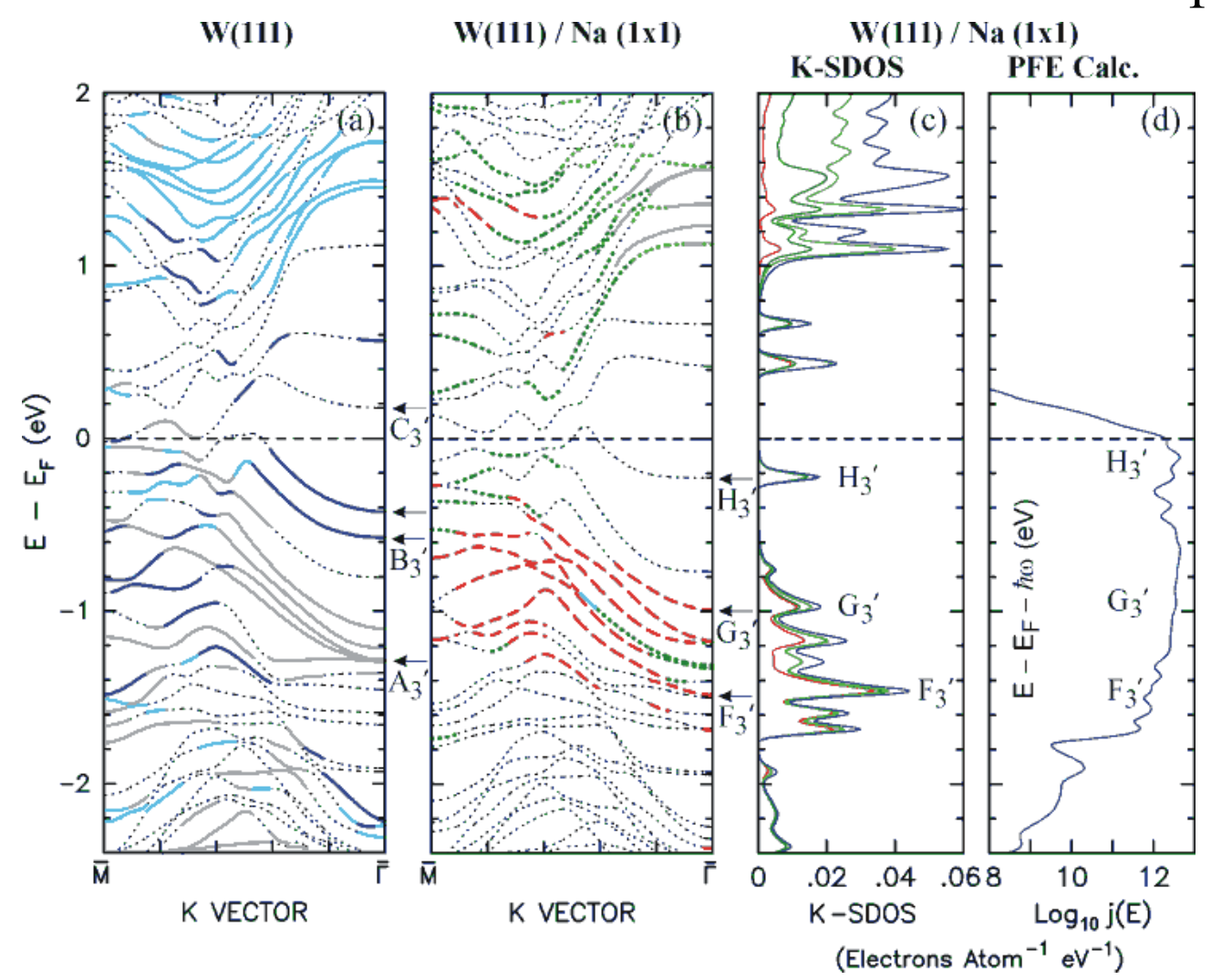

Fig. 12. (Color online) Dispersion plots along $\bar{M}\bar{\Gamma}$ of (a) clean W(111) and (b) W(111)/Na (1×1) plotted in the surface Brillouin zone of the overlayer. Surface states and surface resonances are shown by thick curves. The predominant symmetry in the surface layer is shown by the line style [*s*, red (dashed); *p*, green (dotted); and *d*, blue or grey (solid)]. Bulk and intermediate states are shown by thin grey dotted lines. (c) K-SDOS of W(111)/Na (1×1). The successive curves in the cumulative plots show the contributions *s*-like (red), $p_x$+$p_y$-like (dark green), $p_z$-like (light green) and *d*-like (blue) states, respectively. (d) Calculated TED in PFE for W(111)/Na (1×1) with 3.05 eV photons, based on surface photoexcitation and plotted as a function of the initial state energy.

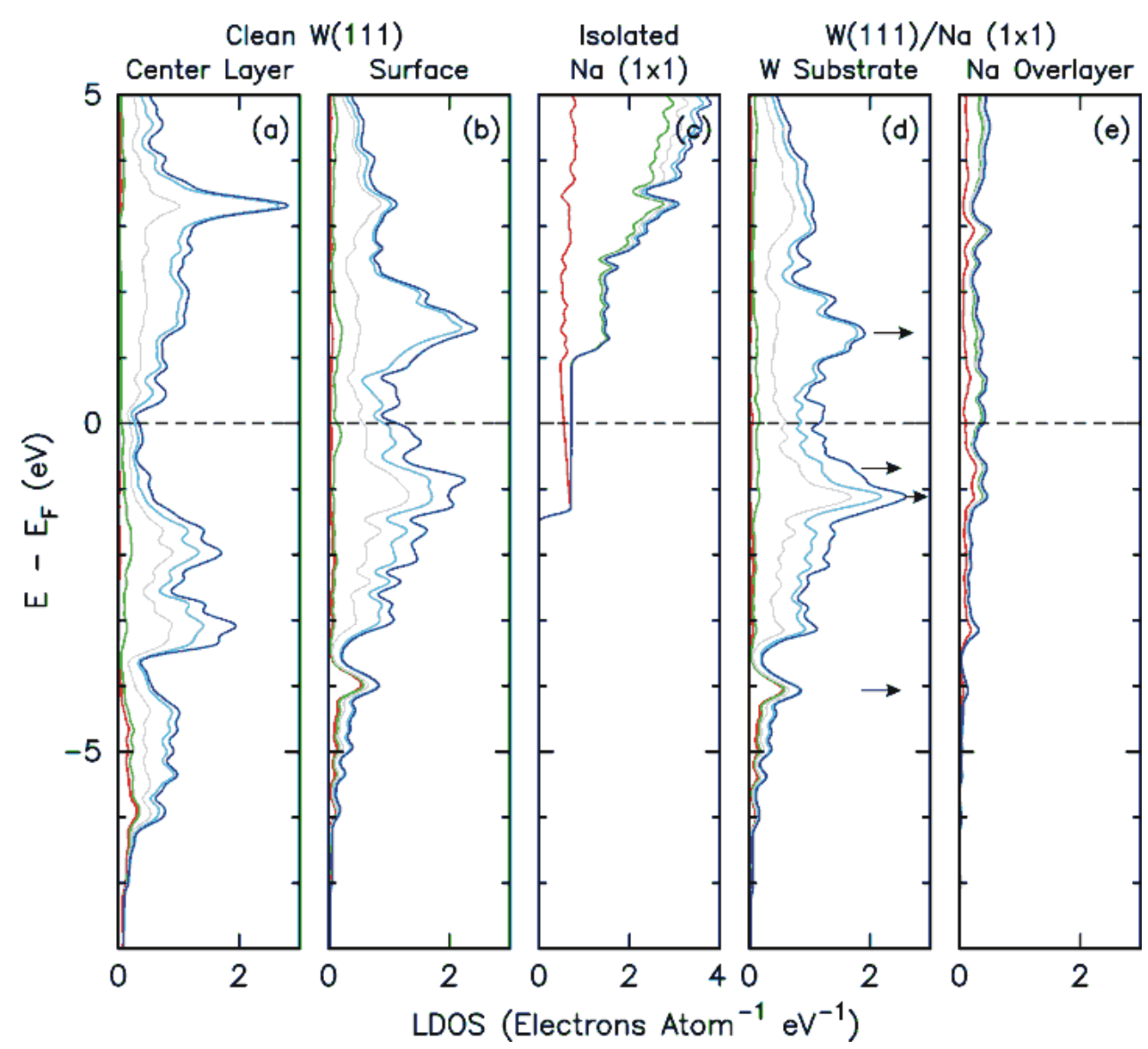

Fig. 13. (Color online) LDOS of clean W(111) in the (a) central (bulk) layer and (b) surface layer. (c) LDOS in an isolated Na (1×1) layer. LDOS of W(111)/Na (1×1) in the (d) W substrate and (e) Na overlayer. In these cumulative plots, the areas between successive curves show the contributions of *s*-like (red), *p*-like (green), $d_{xy}$+$d_{x^2-y^2}$-like (grey), $d_{xz}$+$d_{yz}$-like (light blue) and $d_z^2$-like (dark blue) states, respectively.

Table III. Comparison between the energies of the peaks observed in the TEDs in FE and PFE from clean W(111) ($A_3$ to $C_3$)[16,17] and the energies of the calculated peaks of clean W(111). Also listed are the energies and symmetries of the calculated peaks in FE from W(111)/Na (1×1) ($F_3'$ to $H_3'$). *s.r.* stands for surface resonances and *int.* for intermediate states (states having both surface and bulk character).

| Experimental | | Calculated | | | |
|---|---|---|---|---|---|
| Peak Label | $E–E_F$ (eV) | Peak Label | $E–E_F$ (eV) | Character | Symmetry in Overlayer [Substrate] |
| $A_3$ | −1.4(2)[20] | $A_3'$ | −1.30(2) | *s.r.* | [$d$] |
| $B_3$ | −0.65(5)<br>−0.75(5)[20] | $B_3'$ | −0.57(2),<br>−0.42(2) | *s.r.* | [$d_z^2$] |
| $C_3$ | +0.2(1) | $C_3'$ | +0.18(2) | *int.* | [$d_z^2$] |
| | | $F_3'$ | −1.45(2) | *s.r.* | $s$ [$d_z^2$] |
| | | $G_3'$ | −0.92(2) | *s.r.* | $s$ [$d_z^2$] |
| | | $H_3'$ | −0.22(2) | *int.* | $s$ [$d_z^2$] |

### V.2. Layer densities of states of W(111)/Na (1×1)

When comparing the LDOS in the central (bulk) layer of clean W(111) (Fig. 13a) with that in the surface layer (Fig. 13b) a number of surface states and surface resonances are identified at about −4.0 eV, −1.0 eV, −0.3 eV and +1.4 eV (Fig. 13b). The LDOS of an isolated Na layer calculated using a supercell in which Na atoms occupy the same sites as in the W(111)/Na (1×1) supercell and the W sites are empty is shown in Fig.13c. The magnitude of the LDOS is constant in the range from −1.4 eV to + 1.0 eV, with some *s*-like component being increasingly replaced by a $p_x$+$p_y$-like component. Above +1.0 eV an additional strong $p_z$-like component is observed. When the Na (1×1) overlayer is adsorbed on the clean W(111) substrate the LDOS of the overlayer (Fig. 13e) is greatly modified and spreads out in energy to about −4.3 eV and shows several strong peaks. The surface resonance peaks in the LDOS of the substrate at −4.1eV, −1.1 eV, −0.7 eV and +1.4 eV are slightly modified and extend into the substrate.

## VI. CONCLUSIONS

We interpret peaks observed in the TEDs in FE and PFE from W(100) with a Na c(2×2) overlayer and from W(110) with a Na (2×2) overlayer[15] by calculating the electronic structure and the total energy distributions of the FE and PFE currents from these interfaces. We show that Na states hybridize with surface states and surface resonances of clean W, shifting their energies, and resulting in various Na-induced peaks in the observed TEDs in FE and PFE that are consistent with experiment. We find that the strong Swanson hump $B_1$ in the TEDs in FE and PFE from clean W(100) is suppressed due to Na adsorption. Instead a new peak $F_1$ is observed in the TEDs in FE and PFE from W(100)/Na c(2×2) due to hybridization of the *s*-like valence states of the Na c(2×2) overlayer with the $d_z^2$-like surface states of the Swanson hump at $\bar{\Gamma}'$. The surface states of W(100) shift thereby by −1.2 eV.

The different symmetry of the W(100)/Na c(2×2) overlayer with respect to that of the clean substrate is responsible for folding back the high symmetry point $\bar{M}$ in the SBZ of the W(100) substrate to $\bar{\Gamma}'$ in the SBZ of the overlayer, allowing states at $\bar{M}$ in the SBZ of the substrate to contribute to FE and PFE. $p_z$-like conduction states of the Na c(2×2) overlayer hybridize with $d_z^2$-like surface resonances of the W(100) substrate that have been folded back from $\bar{M}$ to $\bar{\Gamma}'$, shifting the surface resonances by about −0.5 eV. Bulk photoexcitation to these states yields peak $J_1$ observed in the TED in FE from W(100)/Na c(2×2). A similar effect is observed when the symmetry point $\bar{S}$ in the SBZ of the W(110) substrate is folded back to $\bar{\Gamma}'$ in the SBZ of the (2×2) overlayer. Bulk photoexcitation involving these W states that hybridize with the Na states is responsible for the strong peak $K_2$ observed in the TED in PFE from W(110)/Na ($S_{1/4}$).

When a Na c(2×2) overlayer is adsorbed on W(100), the charge is redistributed among the angular momentum states, modifying the spatial distribution of charge in the vicinity of the surface. We show that the net effect is a strong outwardly-directed dipole layer resulting in a decrease in the work function. The work function we calculated for W(100)/Na c(2×2) is in good agreement with the experimentally observed work function at 1 ML coverage.

Contrary to Ishida's results[8], our linear augmented plane wave calculations show clearly that with increased Na coverage from $\theta = 1/6$ to $\theta = 2/5$ the strong *s*-like peak below $E_F$ and the strong *s* and $p_z$-like peak above $E_F$ in the SDOS of W(110)/Na that are dominated by states at $\bar{\Gamma}'$ shift by −0.4 eV and −1.3 eV respectively. This energy shift is due to the increased intralayer Na interactions with increased coverage that broaden the Na states resulting in the shift of the states at $\bar{\Gamma}'$ to lower energy.

## ACKNOWLEDGMENTS

This work was supported by a Natural Sciences and Engineering Research Council of Canada (NSERC) Discovery Grant. The author Z. Ibrahim wishes to thank Dr. A.I. Shkrebtii for helpful discussions during the final stages in the paper preparation.

## REFERENCES

* Corresponding author; zahraa.ibrahim@uoit.ca